\documentclass[reprint,
 amsmath,amssymb,
 aps,superscriptaddress
]{revtex4-2}

\usepackage{graphicx}
\usepackage{dcolumn}
\usepackage{bm}
\usepackage{url}

\usepackage[utf8]{inputenc}
\usepackage[LGR,T1]{fontenc}
\usepackage{hyperref}

\newcommand\Koppa{\begingroup\fontencoding{LGR}\selectfont\char19\endgroup}

\begin{document}

\preprint{}

\title{Scintillation response of  cryogenic CsI to few-keV and sub-keV nuclear recoils }

\author{J.I.\ Collar}
\email{collar@uchicago.edu}
\affiliation{Enrico Fermi Institute, Kavli Institute for Cosmological Physics, and Department of Physics\\
University of Chicago, Chicago, Illinois 60637, USA}

\affiliation{Donostia International Physics Center (DIPC), Paseo Manuel Lardizabal 4, 20018 Donostia-San Sebastian, Spain}

\affiliation{Ikerbasque, Basque Foundation for Science, Plaza Euskadi 5, 48013, Bilbao, Spain}

\author{C.M.\ Lewis}
\email{mark.lewis@dipc.org}
\affiliation{Enrico Fermi Institute, Kavli Institute for Cosmological Physics, and Department of Physics\\
University of Chicago, Chicago, Illinois 60637, USA}

\affiliation{Donostia International Physics Center (DIPC), Paseo Manuel Lardizabal 4, 20018 Donostia-San Sebastian, Spain}

\author{A.\ Simón}
\email{ander.simon@ific.uv.es}
\affiliation{Enrico Fermi Institute, Kavli Institute for Cosmological Physics, and Department of Physics\\
University of Chicago, Chicago, Illinois 60637, USA}
\affiliation{Donostia International Physics Center (DIPC), Paseo Manuel Lardizabal 4, 20018 Donostia-San Sebastian, Spain}
\affiliation{Instituto de Física Corpuscular, CSIC \& Universitat de València,
Calle Catedrático José Beltrán 2, 46980, Paterna, Spain}

\author{S.G.\ Yoon}
\email{sgyoon@uchicago.edu}
\affiliation{Enrico Fermi Institute, Kavli Institute for Cosmological Physics, and Department of Physics\\
University of Chicago, Chicago, Illinois 60637, USA}



\date{\today}

\begin{abstract}
Monochromatic neutron emissions from photonuclear sources $^{88}$Y/Be and $^{124}$Sb/Be  are employed to obtain the response of pure (undoped) cesium iodide  at 80 K. The use of a low-noise, high-quantum-efficiency avalanche photodiode in combination with a novel waveshifter results in a 70 eV analysis threshold. This reach allows to observe signals from sub-keV nuclear recoils originating in neutron scattering. The extracted quenching factor drops much faster towards low energy than the extrapolation of a model developed for room-temperature CsI[Na]. We comment on the impact of our measurement on  planned use of cryogenic CsI in neutrino physics and dark matter experiments. 
\end{abstract}

\maketitle


\section{Introduction} Recent times have seen an emphasis on the identification and characterization of radiation detection media capable of sensing modest energy depositions, specifically those from nuclear recoils induced by the scattering of hypothetical Weakly Interacting Massive Particles (WIMPs) or by low-energy neutrinos. While the quest for WIMPs continues, the recent observation of Coherent Elastic Neutrino-Nucleus Scattering (CE$\nu$NS) \cite{science} has further validated the interest in this area.

Close to two decades of dedicated research at the University of Chicago led to the identification of sodium-doped cesium iodide (CsI[Na]) as the medium with ideal characteristics to permit the first CE$\nu$NS measurement. Its multiple advantages for this application are described in \cite{NIMcenns,nicole,bjorn}. Pure (i.e., undoped) CsI operated at cryogenic temperature has since been  recognized as an alternative  sharing those benefits \cite{ESS,csiqf,clovers,kims,chireactor,cohcsi} while providing a light yield near the theoretical maximum for inorganic scintillators (where this yield is fully determined by the number of electron-hole pairs along the ionization track \cite{DORENBOS2002208}). This could reduce the demonstrated $\approx$5 keVnr energy threshold of CsI[Na] for CE$\nu$NS detection \cite{science,NIMcenns,nicole,bjorn} down to $\approx$1 keVnr \cite{ESS,csiqf} (``nr'' refers to energy carried by the nuclear recoil, as opposed to detected energy, denoted here by ``ee'' for ``electron equivalent''). A first characterization of the response of cryogenic CsI to  6-25 keVnr recoils \cite{csiqf} indicated a promising  behavior identical to that found for room-temperature  CsI[Na] in this energy interval \cite{csinaqf}.

The characterization of a material-specific ``quenching factor” (QF, the fraction of nuclear recoil energy deposited in a detectable form) is of  utmost importance in the context of CE$\nu$NS, as new physics and a misinterpreted QF can be easily mistaken for each other \cite{csinaqf}. The response to sub-keV recoils is hard to predict via theoretical models and can vary across  materials. Several approaches to QF characterization in this difficult-to-reach energy regime have been proposed \cite{jk2,ksu,myprl,sbbefe,scandium,crab,PhysRevD.107.076026,PhysRevD.105.083014}.

 \begin{figure}[!htbp]
\includegraphics[width=1. \linewidth]{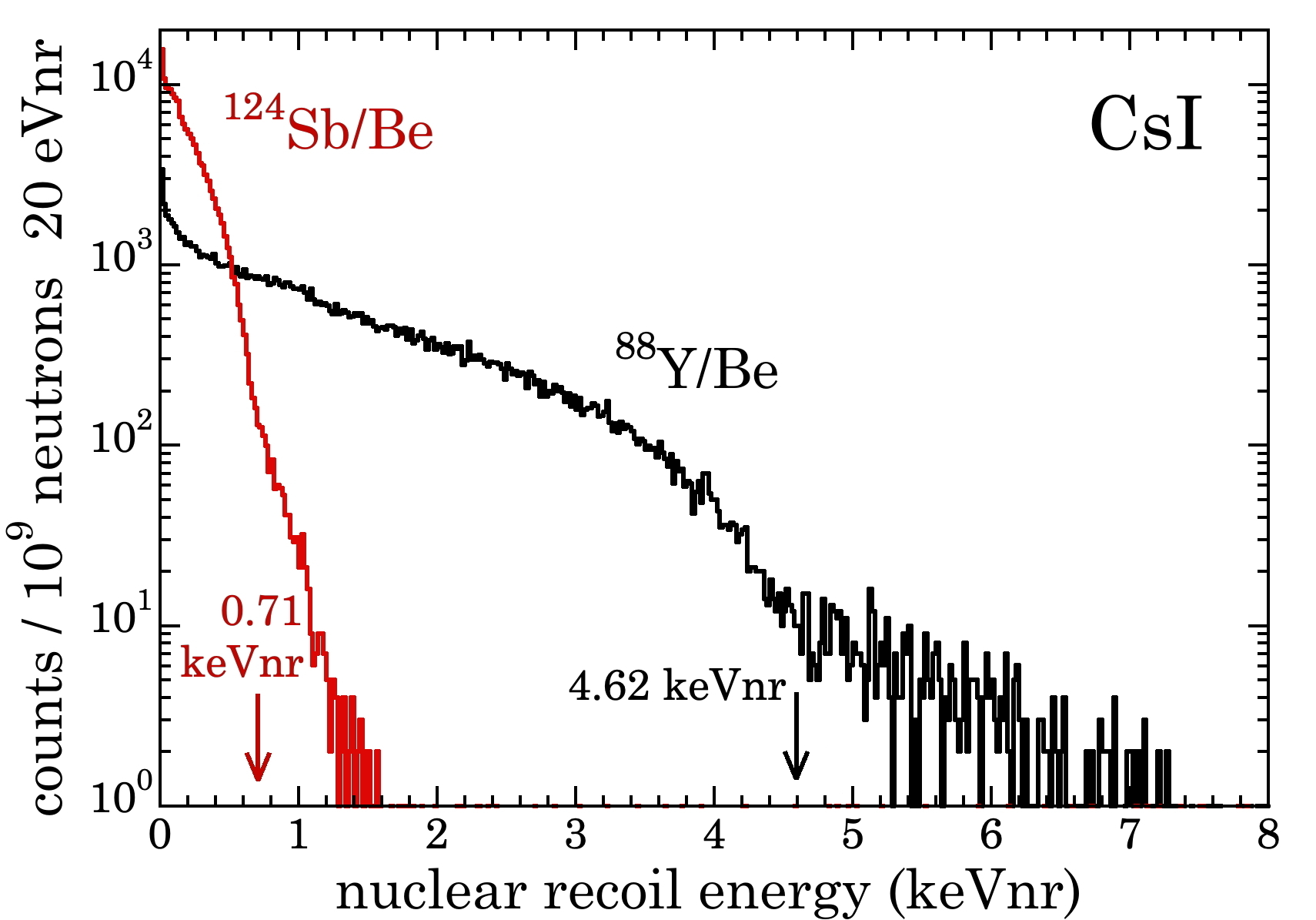}
\caption{\label{fig:epsart}  Simulated nuclear recoil distributions produced by the dominant neutron emission from $^{88}$Y/Be and $^{124}$Sb/Be in the  geometry of Fig.\ 2. Cs and I recoils are indistinguishable due to their similar nuclear mass \cite{NIMcenns}. Vertical arrows indicate the maximum recoil energy from  single scatters. A tail beyond originates in the modest fraction of multiple scattering by neutrons in this small crystal ($\approx$10.7 \% for both sources). }
\end{figure}

Of these, the most convenient is the use of photonuclear radioactive sources that generate monochromatic neutrons capable of inducing a predictable spectrum of low-energy nuclear recoils \cite{myprl} (Fig.\ 1). These have been used to characterize NaI[Tl] \cite{myprl}, superheated liquids \cite{myprl,pico,eric}, silicon \cite{siqf}, germanium \cite{geqf1,geqf2,PhysRevD.105.122002} and most recently to calibrate the XENON-nT and LUX-ZEPLIN detectors prior to their first observation of solar neutrinos via CE$\nu$NS \cite{xenon1,xenon2,LZ1,LZ2}.  Here we exploit this approach to study the QF of cryogenic CsI below 4.6 keVnr, extending our previous work in \cite{csiqf} towards lower energy. Signals from sub-keV nuclear recoils are unequivocally detected. To our knowledge, this is a first instance for inorganic scintillators read out exclusively via their light output. 

In what follows, we describe the details of the methodology leading to this  sensitivity, as well as the analysis steps taken towards a characterization of the QF in this energy region. We finish by offering a commentary on the impact of our findings on the prospects for use of cryogenic CsI in this area of research.

\begin{figure}[!htbp]
\includegraphics[width=1. \linewidth]{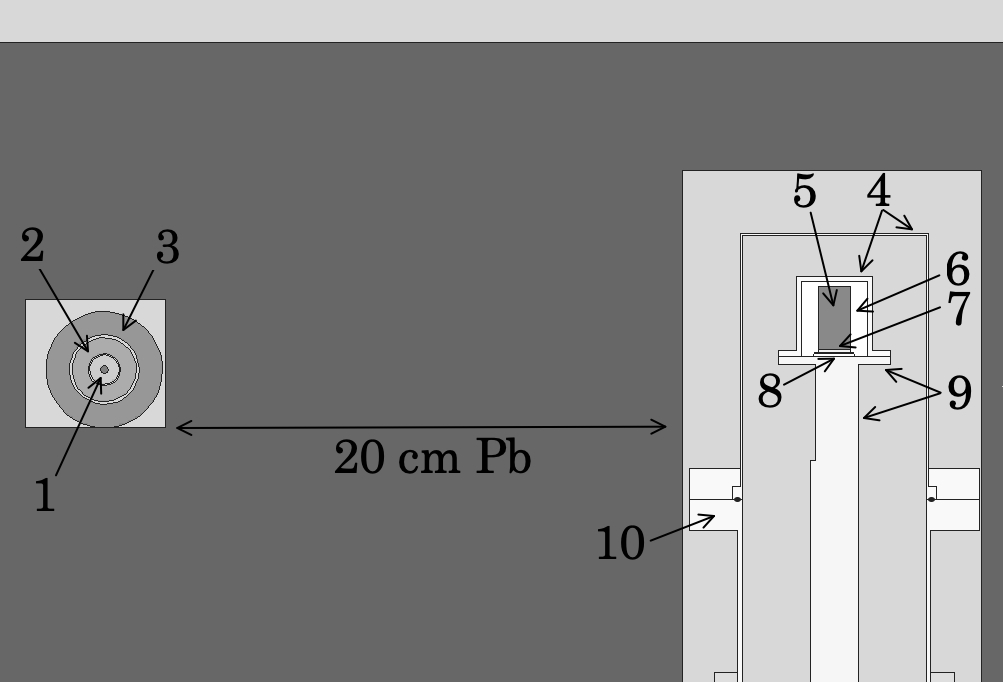}
\caption{\label{fig:epsart} Elements of the $^{88}$Y irradiation geometry: 1) V-vial containing  evaporated  $^{88}$Y source, 2)  stainless steel (SS) container, 3) beryllium oxide ceramic or aluminum metal, 4) aluminum endcap and infrared shield, 5) PTFE-wrapped cesium iodide crystal, 6) holder (sponge), 7) acrylic window coated with NOL-9 waveshifter, 8) LAAPD on alumina substrate, 9) copper cold finger and cold plate, 10) SS cryostat flange. $^{124}$Sb irradiations varied only in internal source geometry.}
\end{figure}

\section{Experimental approach} 5 mCi of $^{88}$Y in liquid solution were obtained from Los Alamos National Laboratory through \cite{nidc}. Its evaporated, triple-sealed residue was surrounded by a beryllium oxide ceramic shell \cite{myprl} able to convert gammas emitted above the 1.66 MeV two-body dissociation threshold for $^{9}$Be($\gamma$,n)$^{8}$Be  \cite{alan1} into monochromatic neutrons. For $^{88}$Y/Be,  a dominant neutron energy is emitted at 152.3 keV \cite{alan1}. Separately, 10 mCi of $^{124}$Sb activity were produced via thermal neutron activation of high-purity antimony pellets \cite{ESPI} at the Oregon State University TRIGA reactor. Surrounded by the same BeO container, neutron yields at 23.5 keV and 379 keV are expected \cite{alan1}. The second is  just a small fraction of the first, somewhere in the interval 2-4\%  \cite{alan1,hunt}. For present intents and for both sources, anisotropic deviations from monochromaticity \cite{hunt} are negligible.  

The neutron yield of each source was measured using a $^{3}$He counter \cite{lnd} surrounded by 5 cm of high-density polyethylene moderator and an external 0.6 mm cadmium metal layer. For Y/Be, the initial intensity was 3,860 n/s. This was 4,570 n/s for Sb/Be. We assign a 5\% uncertainty to these yields, derived from our previous study of neutron sources of known activity using this same detector assembly (see ``Cd-wrapped" in Table B.1 of \cite{drew}). The decay of the sources (T$_{1/2}=$ 106.6 days for $^{88}$Y and T$_{1/2}=$ 60.2 days for $^{124}$Sb) was taken into account to extract the average activity during data-taking runs described below (a reduction in yield to 3,410 n/s and 3,160 n/s, respectively). 

Fig.\ 2 shows the positioning of these sources with respect to a vacuum cryostat housing a 1.27 cm  diameter, 2.54 cm long cylindrical crystal of pure CsI.  The crystal was procured from AMCRYS stock \cite{proteus}, same origin as the larger sample in \cite{csiqf}. The cold finger of the cryostat was immersed into a liquid nitrogen Dewar, producing  a measured 80 K temperature at the crystal. Twenty centimeters of lead between source and detector are used to shield the majority of gamma emissions from the source, while only minimally moderating neutrons \cite{myprl}. Data-taking consisted of alternating daily exposures, first using BeO, then with this ceramic replaced by a shell of aluminum metal of identical dimensions, inert as neutron emitter. We refer to each pair of measurements as a ``run''. A total of five to six runs were taken with each source, over a period of one month. Due to the very similar macroscopic cross section for gammas in BeO and Al at relevant energies \cite{noteref}, the exposure-normalized difference between scintillation spectra (Y/Be-Y/Al and Sb/Be-Sb/Al residuals) contains energy depositions by neutrons only \cite{myprl}.

\begin{figure}[!htbp]
\includegraphics[width=1. \linewidth]{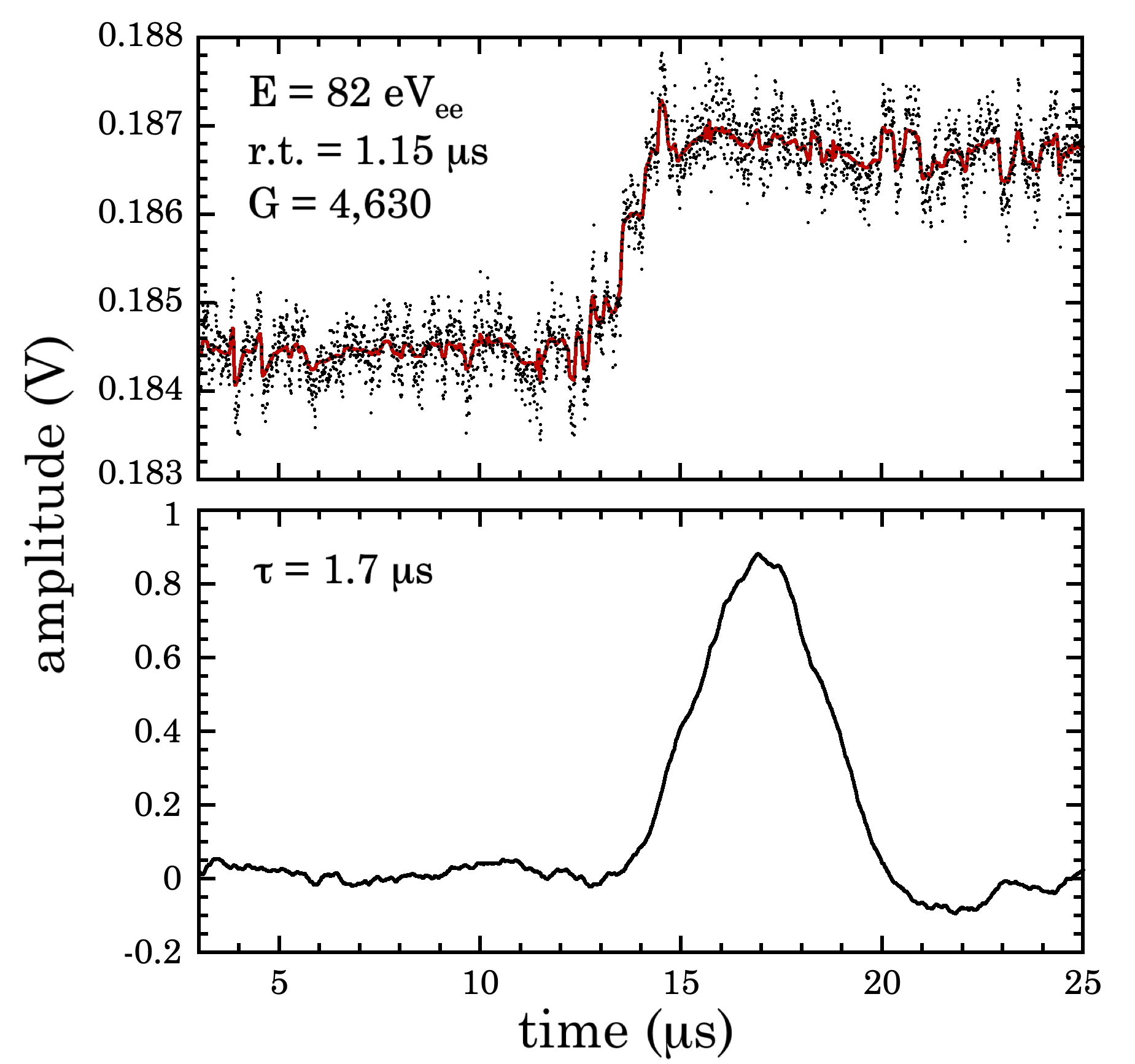}
\caption{\label{fig:epsart} {\it Top:}  preamplifier trace from an 82 eV scintillation signal in the CsI crystal. Fitted rise time and LAAPD gain are indicated. {\it Bottom:} the same signal  digitally shaped with the integration time shown. Using a light yield of 106 photons/keV at 80 K \cite{csiqf} and a  70\% LAAPD quantum efficiency (see text), this  corresponds to the detection of $\approx$6 photons and a trigger threshold at 4 photons, similar to that  in \cite{rmd2}. }
\end{figure}

The PTFE-wrapped scintillator was monitored by a 1.3$\times$1.3 cm$^{2}$  silicon large-area avalanche photodiode (LAAPD) obtained from \cite{rmd}. LAAPDs show an excellent low-noise performance at liquid nitrogen temperature and maximum internal gains of $\approx$10,000, permitting light detection at few-photon sensitivity for cryogenic sensor areas of up to 45 cm$^{2}$ \cite{rmd2}. As  in \cite{ESS}, a thin 1 mm acrylic window coated by NOL-9 waveshifter \cite{nol1,nol2} was placed at the exit window of the crystal, coupled with optical grease to the LAAPD. This results in an expected quantum efficiency for photon detection of $\approx$70\% at the 588 nm wavelength of NOL-9 emission \cite{ESS}.

Two evaporated $^{55}$Fe sources were placed inside the cryostat, one illuminating the surface of the CsI crystal opposite to the LAAPD, the other in direct contact to the unencumbered LAAPD surface.  In the first case the 5.9 keV $^{55}$Fe x-ray was used to provide a scintillation energy reference, with assumed proportionality for lower-amplitude signals (zero energy at zero amplitude). In the second, to calculate the internal LAAPD gain (Fig.\ 3, top panel) through the average energy for electron-hole pair production in 80 K silicon (3.7 eV \cite{4326778}) and the known gain of the charge preamplifier used for readout (Cremat CR-110-R2, 0.7 volts/pC into 50 $\Omega$). CsI scintillation signals and those from radiation interactions with the LAAPD can be easily distinguished by their preamplifier rise time (Figs.\ 3,4). This corresponds to the $\approx$1 $\mu$s scintillation decay constant of CsI at 80 K \cite{csiqf,decay} and to the $\approx$100 ns electron transport time in a $\approx$220 $\mu$m-thick LAAPD, respectively (Fig.\ 4).

\begin{figure}[!htbp]
\includegraphics[width=1. \linewidth]{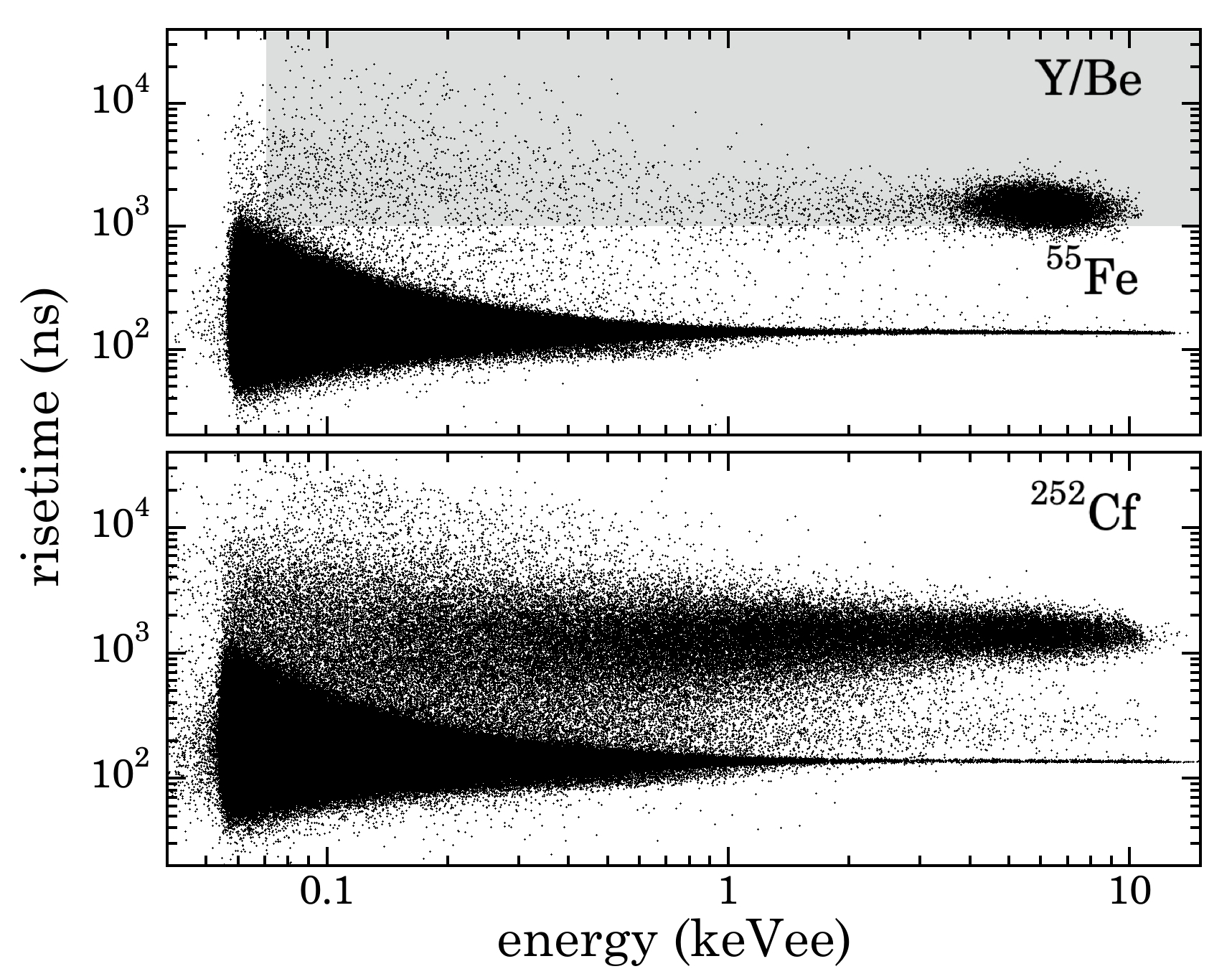}
\caption{\label{fig:epsart}{\it Top:} preamplifier rise time of scintillation ($\approx$1 $\mu$s) and LAAPD ($\approx$100 ns) signals during a Y/Be run. The 5.9 keV $^{55}$Fe deposition is visible for the first. The rise time of a charge-trapping noise noticeable below 1 keV (scintillation) is broadened by the effect of electronic noise  (Fig.\ 3). A grayed region indicates events accepted for analysis (see text). {\it Bottom:} same during a $^{252}$Cf calibration run (see text).  }
\end{figure}

Data acquisition was performed by programming a triangular-shaping algorithm on the FPGA front-end (NI PXIe-7966R) of a NI 5734 digitizer, set to sample the AC-coupled preamplifier output at 120 GS/s. A  trigger threshold at a scintillation energy of 55 eV  was imposed on the amplitude of the shaped signal, sufficiently away from baseline noise fluctuations (Fig.\ 3, bottom panel). The triggering efficiency was monitored several times during data-taking using a programmable pulser set to produce rise-times similar to scintillation signals. An excellent trigger stability was observed, with constant 100\% pulser signal acceptance in the $>$70 eV energy region chosen for analysis (Figs.\ 4,5). 

 Contrary to our approach in \cite{csiqf} and aiming to minimize sources of stray electronic noise, no active temperature control of the crystal was attempted. The gain of LAAPDs operated close to the Geiger regime is sensitive to temperature fluctuations \cite{temp1,temp2}. Changes in laboratory ambient temperature  during the weeks-long data taking period induced gain drifts readily visible by monitoring the $^{55}$Fe scintillation reference. While relatively modest (2\% rms, maximum of 8.9\%), these were controlled by dividing the data into 2 hour periods -shorter than the fastest deviation observed- and gain-shifting the data according to the $^{55}$Fe beacon. A perfect cancellation of the 5.9 keV signal in the Be-Al residuals from all runs (Fig.\ 5) evidenced the effectiveness of this correction.

\begin{figure}[!htbp]
\includegraphics[width=1. \linewidth]{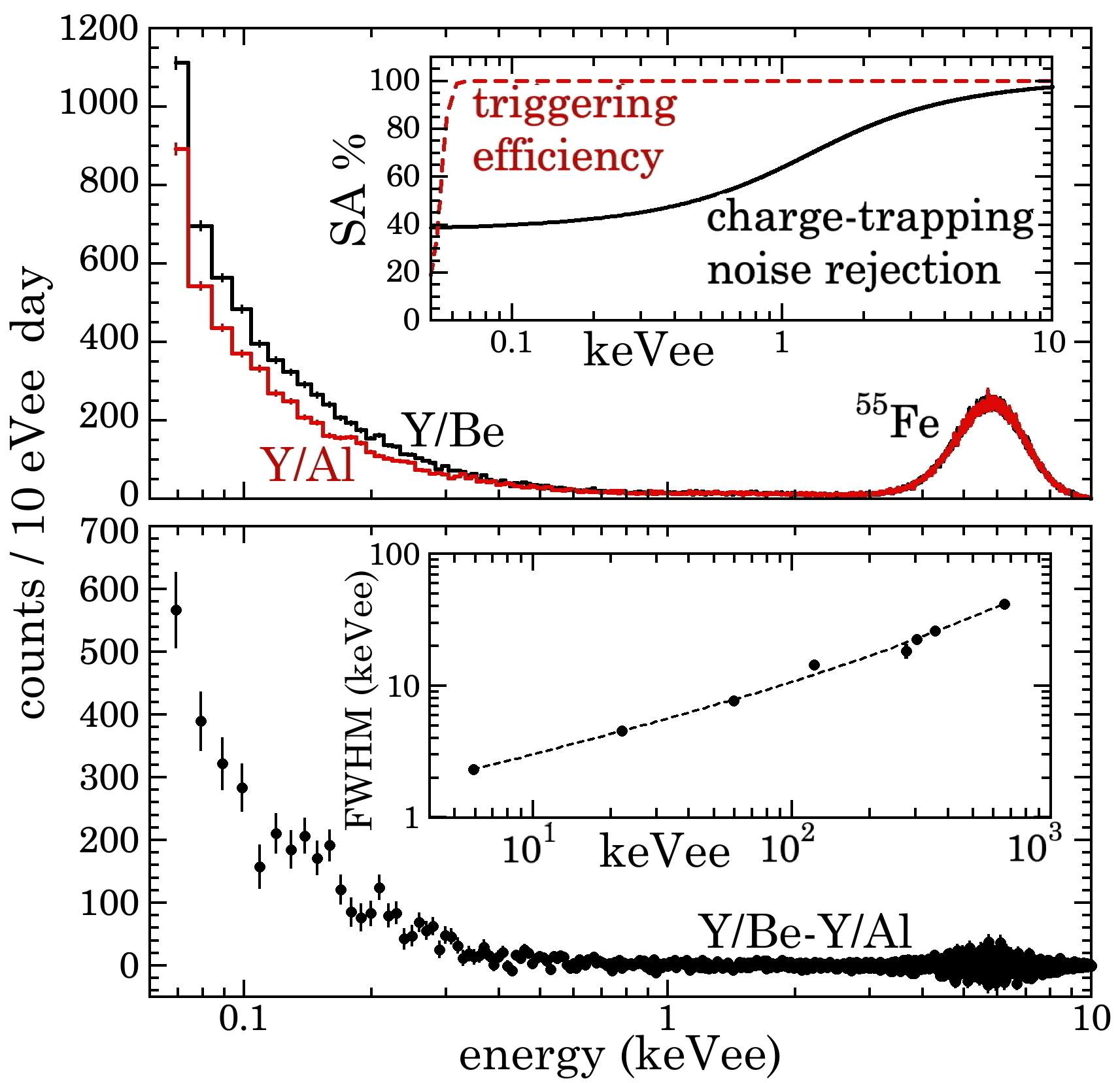}
\caption{\label{fig:epsart}{\it Top:} Cumulative spectra from Y/Be and Y/Al runs following the rise-time acceptance cut and normalization to the same exposure. {\it Inset:} signal acceptance (SA) for pulser calibration events and scintillation signals in CsI (see text). {\it Bottom:} Y/Be-Y/Al residual, corrected for SA. The excess due to neutron scattering  is visible (see text). Error bars are statistical. {\it Inset:} energy resolution measurements and their fit (see text). Most error bars are encumbered by  datapoints.}
\end{figure}

 The manufacture of LAAPDs typically involves use of low-resistivity NTD (neutron transmutation doped) silicon wafers. Although a thermal annealing process follows high-flux thermal neutron irradiation of NTD silicon boules in a nuclear reactor, the defects and lattice dislocations that survive can act as charge-trapping centers. We have observed evidence for this process by briefly exposing isolated LAAPDs to high-intensity radiation sources (neutron, gamma) and noticing an induced prompt surge in fast-rise-time  low-energy signals, vanishing with a relaxation time of $\approx$1 hour (at 80 K) once the source is removed. Such signals appeared at a constant few-Hz rate during present runs (Fig.\ 4). This nuisance can be efficiently removed by imposing a cut  accepting signals only above 1 $\mu$s rise time (Fig.\ 4), at the expense of a diminished scintillation signal acceptance (SA) near threshold. This SA was measured by exposing the  detector to a $^{252}$Cf source located outside of its lead shield. Penetrating neutrons create a continuum of nuclear recoil signals below few keVee (Fig.\ 4). The rise-time distributions of scintillation and LAAPD signals were fitted by Gaussians, as a function of energy. The SA derived from these fits has negligible uncertainty and is shown in the top inset of Fig.\ 5. 

 The full-width-at-half-maximum (FWHM) energy resolution of the detector was characterized using gamma ray lines from $^{137}$Cs, $^{133}$Ba, $^{57}$Co, and $^{241}$Am sources placed outside of the cryostat, in addition to the internal $^{55}$Fe x-ray. This required substitution of the CR-110 preamplifier by alternatives (CR-111, CR-112) with progressively lower gains, to avoid their saturation. The resolution was referred to the nominal energy of each peak. The energy ($E$) dependence was fitted using a functional form FWHM$= a + b\sqrt{E+cE^{2}}$, customary for scintillators. When both FWHM and $E$ are in units of keV, the best-fit parameters are $a=0.0508\pm0.1476$ keV, $b=0.9197\pm0.0303$ keV$^{1/2}$, and $c=0.0033\pm0.0003$ keV$^{-1}$. As expected from the difficulty in using x-ray sources below 5.9 keV capable of traversing the PTFE reflector around the crystal, the uncertainty in the sub-keV resolution is dominated  by that of the parameter $a$. Data and fit are shown in the bottom inset of Fig.\ 5.

\section{Analysis and QF results} Y/Be-Y/Al and Sb/Be-Sb/Al residuals, once corrected for SA, gain-shifted against temperature fluctuations, and normalized for exposure, display evident near-threshold excesses due to neutron scattering (Figs.\ 5,6). For $^{124}$Sb irradiations this excess appears very close to threshold, but is consistently present in individual runs (Fig.\ 6 inset).

Detailed MCNPX-PoliMi \cite{mcnpx} neutron transport simulations using the geometry in Fig.\ 2 were performed to generate databases containing sufficient nuclear recoil statistics (Fig.\ 1). Improved neutron cross-section libraries \cite{alan2} were involved. We employed two different implementations of the Markov chain Monte Carlo  method (MCMC) to test models of  CsI response against residual data. One relied on a direct implementation of the Metropolis-Hastings algorithm \cite{Hastings,GW10}, the second on a popular MCMC package \cite{mcmc1,mcmc2}. Both generated quantitatively similar results. The number of free parameters in these models was kept to a minimum: assumptions made towards this goal were tested, as described below. 

The MCMC algorithms post-process the databases of simulated recoil energies for each combination of free parameter values in  models under test, comparing the obtained synthetic residuals with those in Fig.\ 6. A normalization of simulated neutron histories to the calculated number of neutrons emitted by the source during runs is applied during this process. This includes the decay of the sources. Best-fit parameter values are obtained from the quality of this comparison.

\begin{figure}[!htbp]
\includegraphics[width=1. \linewidth]{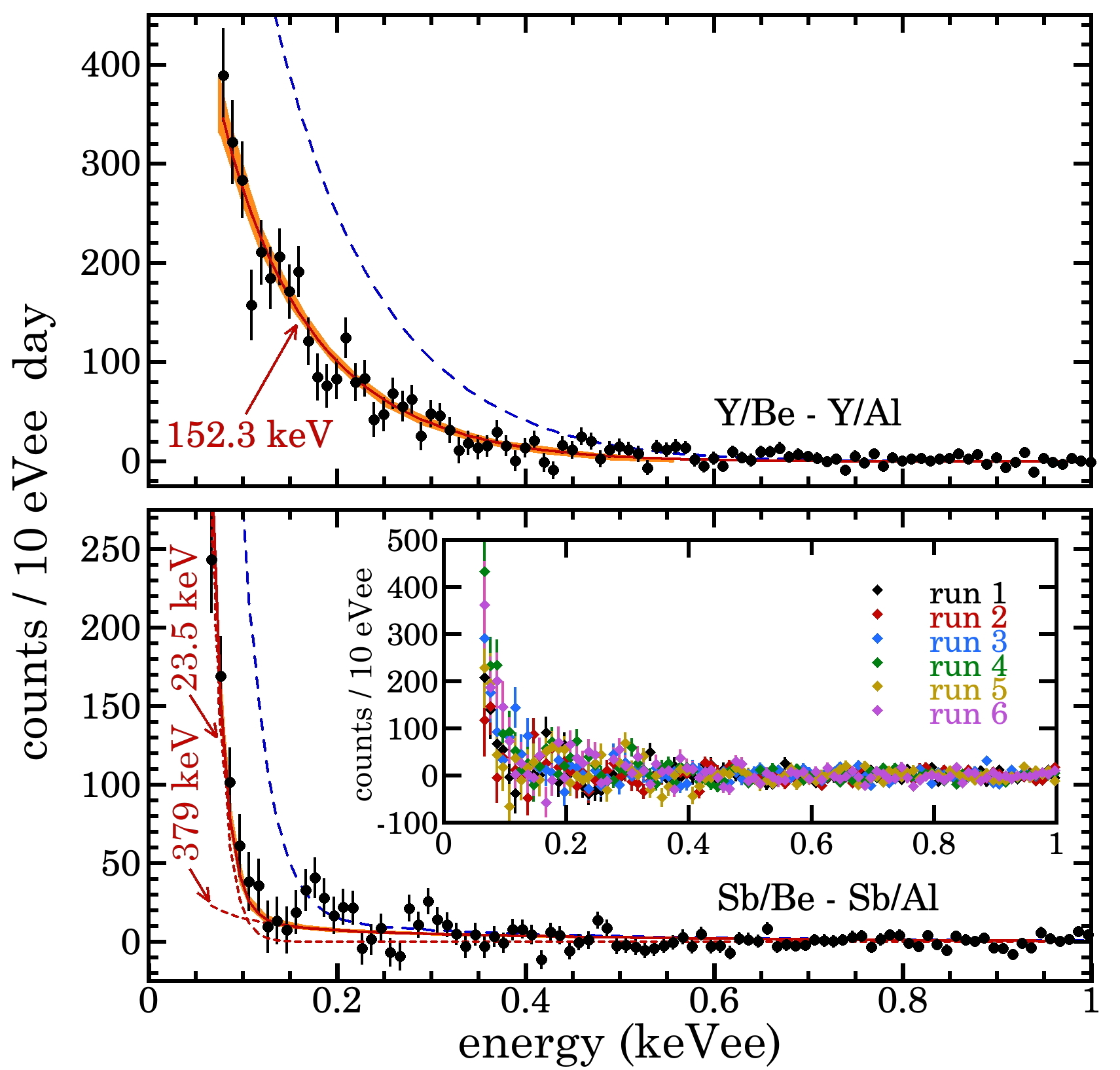}
\caption{\label{fig:epsart} {\it Top:} Y/Be-Y/Al residual (datapoints).  The prediction from the best-fit QF model is shown as a red solid line, with an orange uncertainty band. The energy of the neutrons emitted by the source is indicated. A dashed blue line indicates the incompatible prediction from the model developed for room-temperature CsI[Na] in \cite{csinaqf}, shown in Fig.\ 8. Error bars are statistical. {\it Bottom:} {\it idem} for Sb/Be-Sb/Al. The inset shows the residual from individual runs (see text). }
\end{figure}

To a good approximation, the energy dependence of the QF can be assumed to be linear between two sufficiently-close recoil energies. For $^{88}$Y runs one of those energies was chosen at 4.5 keVnr, close to the maximum for single recoils (Fig.\ 1). The choice of second fulcrum was explored over the interval 1.5-2.5 keVnr, obtaining very similar QF model predictions, but with a mild goodness-of-fit preference for 2.0 keVnr. Values of the QF at these two energies ({\it \Koppa$_{2.0}$, \Koppa$_{4.5}$}) were independently allowed to be  anywhere in the range 0-10 \%, exploring a wide variety of QF behaviors. A third free parameter in the model, denoted by $Y$, represents an overall normalization factor affecting the synthetic residual. It allows for a possible inaccuracy during source activity characterization. This parameter was constrained around $Y=1$ (i.e., perfect agreement with the measured neutron yield) by a Gaussian prior with a width corresponding to the 5\% uncertainty mentioned in the previous section.

For each simulated neutron history the total scintillation energy is computed from individual recoils, typically singles (Fig.\ 1), and the energy-dependent QF under test. This total energy is smeared according to the corresponding resolution before being tallied into the synthetic residual. During a first approach to $^{88}$Y analysis  a fourth free parameter was allowed, representing $a$ in the energy resolution discussion above.  The value preferred by the fit ($a=0.017\substack{+0.016 \\ -0.012}$ keV) is well-within the large uncertainty noted before. A finite value is expected to be favored, as this term comprises all energy-independent factors affecting the resolution (preamplifier and digitizer noise,  leakage current and excess noise factor of the LAAPD). It was also noticed that simply imposing the nominal  $a=0.0508$  only leads to  minor changes (few percent) in the returned {\it \Koppa$_{2.0}$, \Koppa$_{4.5}$}, and $Y$. Following this precautionary cross-check and aiming to minimize the number of free parameters,   the nominal  resolution was adopted henceforth.

\begin{widetext}
\begin{figure*}[!htbp]
\includegraphics[width=1. \linewidth]{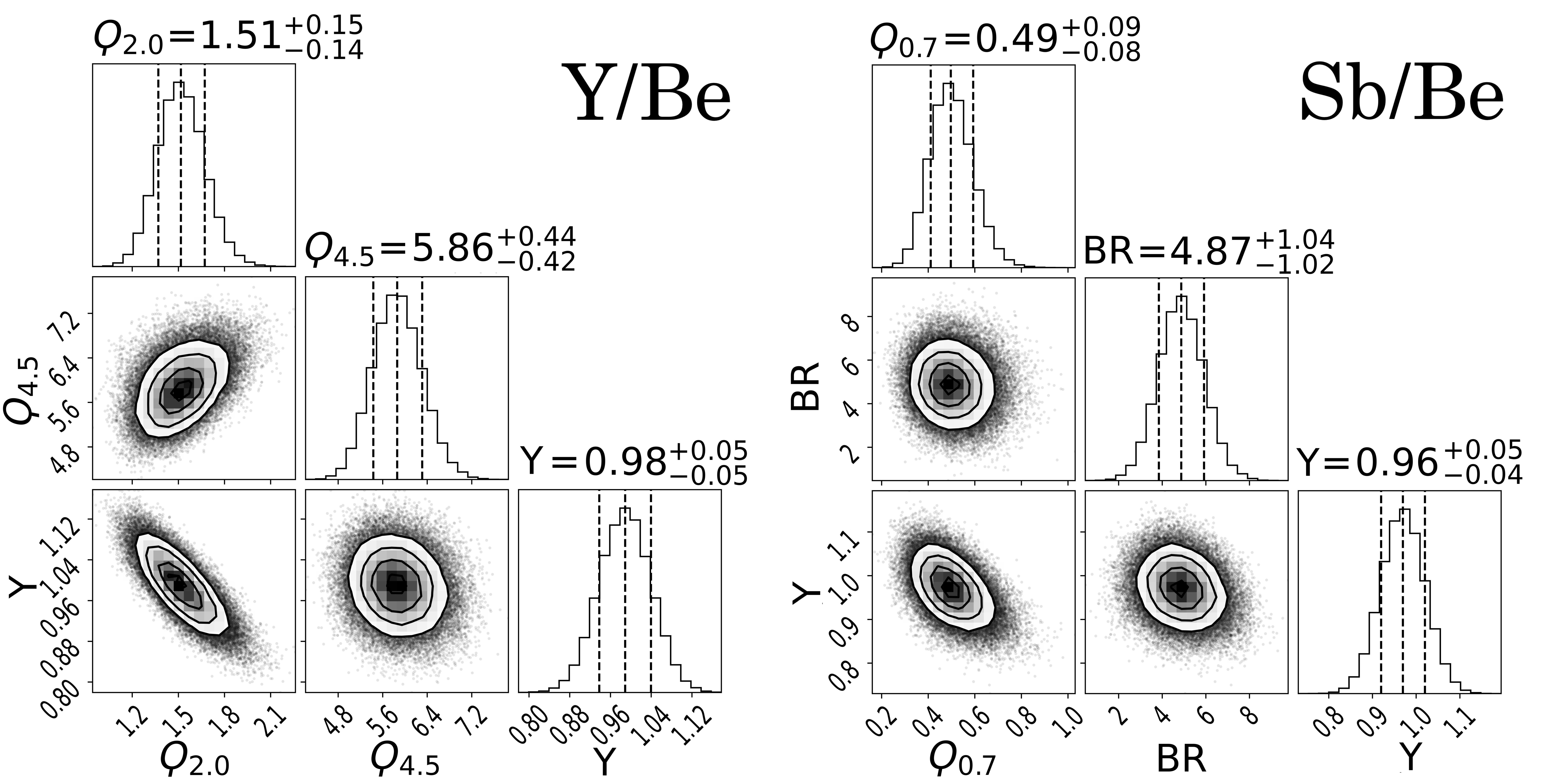}
\caption{\label{fig:wide}MCMC corner plots and best-fit values from the analysis (see text). All parameters except $Y$ are percentages. }
\end{figure*}
\end{widetext}

The same linear QF model was used for a separate  $^{124}$Sb analysis. A similarly-free QF at the endpoint of the recoil distribution ({\it \Koppa$_{0.7}$}) was adopted. The second fulcrum was placed at zero recoil energy. Left as a free parameter, this {\it \Koppa$_{0.0}$} favored a null value. Following this check, a zero QF at zero recoil energy was imposed.

\begin{figure}[!htbp]
\includegraphics[width=1. \linewidth]{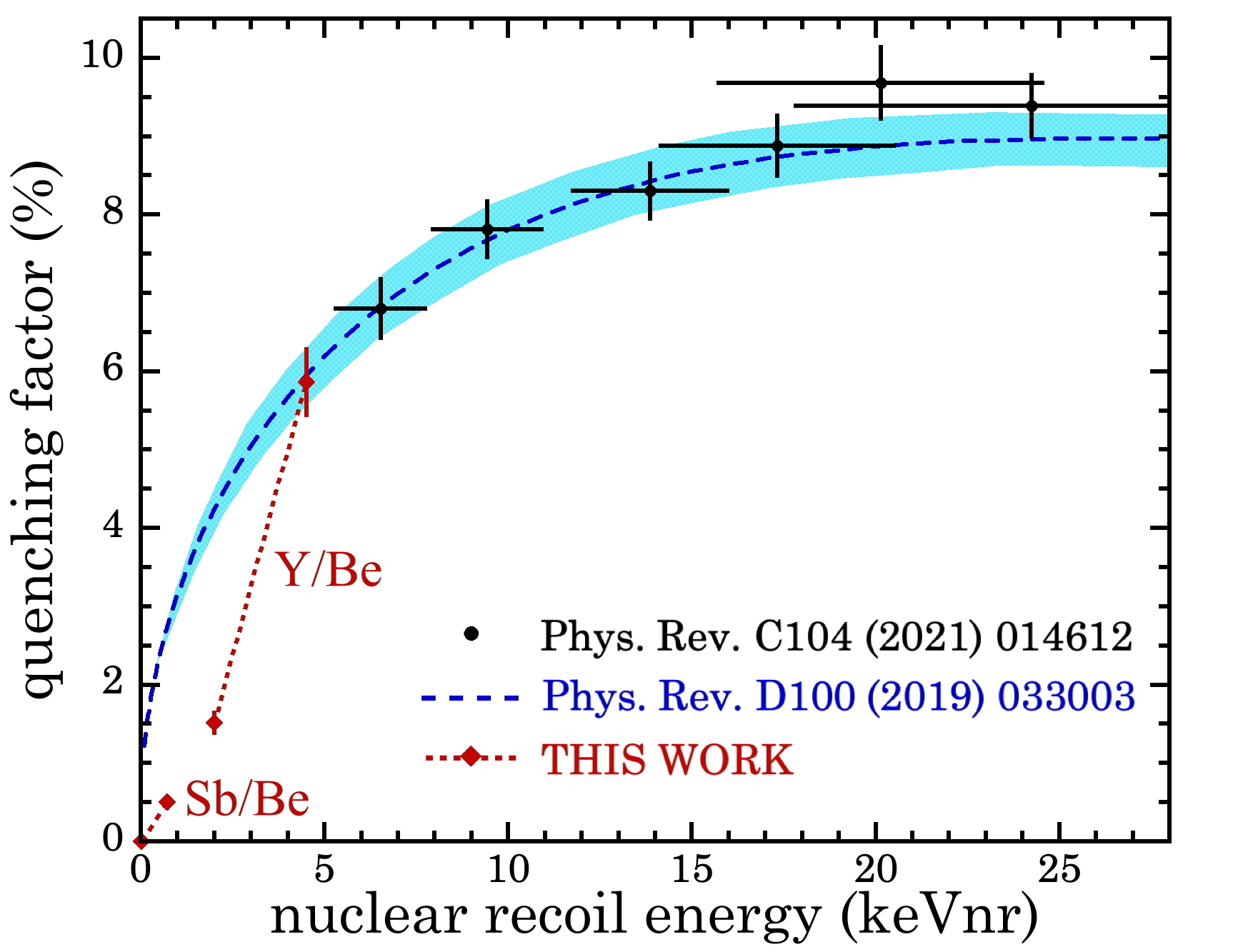}
\caption{\label{fig:epsart} Comparison of present results (80 K) with our  measurements for pure CsI at 108 K \cite{csiqf} (black datapoints). A temperature-independent QF was observed over  \mbox{100-170 K}  in that work.  The QF model for room-temperature CsI[Na] in \cite{csinaqf} is shown by a blue dashed line, with an uncertainty band of the same color (see text). Its extension to low energy adheres to the physics-based model description in \cite{csinaqf}.  }
\end{figure}

A small excess in the experimental Sb/Be-Sb/Al residual noticeable above 0.15 keVee in Fig.\ 6 can be attributed to the subdominant 379 keV neutron emission. Such an excess is also visible in our previous work involving a germanium target exposed to $^{124}$Sb/Be \cite{geqf2}. In view of its presence here, we introduced a free parameter $BR$ to account for the (percent) branching ratio of this secondary emission, taking the dominant 23.5 keV as a reference. A dedicated database of simulated recoils was generated for this second branch. During its post-processing a QF model with expanded energy range was necessary: this included a linear behavior between {\it \Koppa$_{0.7}$} and the value of {\it \Koppa$_{2.0}$} determined by $^{88}$Y analysis, the line between {\it \Koppa$_{2.0}$} and {\it \Koppa$_{4.5}$} found by the same, and above 4.5 keVnr, the  model from \cite{csinaqf} displayed by a blue dashed line in Fig.\ 8. The favored value ($BR=4.87 \substack{+1.04 \\ -1.02}$, Fig.\ 7) is statistically compatible with that from \cite{hunt}   ($3.0\pm0.8$).

Fig.\ 7 displays MCMC corner plots and best-fit values for all parameters involved. A comparison with our previous results for cryogenic (108–165 K) CsI \cite{csiqf} and room-temperature CsI[Na] \cite{csinaqf} is provided by Fig.\ 8.

\section{Commentary} The single QF datapoint in \cite{csiqf,csinaqf} below 3 keVnr is at 2.9 keVnr. It is statistically compatible at a 1.06 $\sigma$ level with present results, which point to a drastic drop in QF for cryogenic CsI below the energies explored in those references and energies involved in the first observation of CE$\nu$NS \cite{science,NIMcenns,nicole,bjorn}. It will be difficult to assess if this same behavior extends to room-temperature CsI[Na] through the use of  photonuclear sources, as $^{88}$Y/Be and $^{124}$Sb/Be would generate signals comprised by at most a few photoelectrons in that medium, even if monitored by a high-quantum efficiency ultra-bialkali \cite{sba} or silicon \cite{Schmailzl_2022} photomultiplier. The room-temperature leakage current of LAAPDs prohibits their use at this  light level. Still, the excellent agreement in QF for both materials above 4 keVnr makes us suspect that present behavior may be a shared feature. {

A question can be entertained that present findings might partially originate in a deviation from the adopted signal proportionality below 5.9 keV, as non-proportionality is common in inorganic scintillators \cite{nonprop1,nonprop2}. No  information on this topic exists  for cryogenic CsI below 5.9 keV \cite{csiqf,mos}, but we observe a strict proportionality in the response to the three lowest calibration energies employed here (59.5 keV, 22.1 keV, 5.9 keV). Sub-keV behavior has been studied for other materials \cite{nai}. In order to bypass this concern  we recommended in \cite{csinaqf} to always relativize the QF  to the energy reference employed during its measurement. In our experience the use of 5.9 keV ($^{55}$Fe, as in this work) or of 59.5 keV ($^{241}$Am, as in \cite{csiqf}) is nearly equivalent for AMCRYS CsI stock (the non-linearity data of Fig.\ 4.5 in \cite{lewis} correct and supersede Fig.\ 10 in \cite{csiqf}).

Our measurement has important implications for planned use of cryogenic CsI, in particular for low-mass dark matter \cite{kims} and reactor  CE$\nu$NS \cite{chireactor}. For example, sensitivity predictions in \cite{chireactor} assume a QF more than an order of magnitude higher than allowed by $^{124}$Sb/Be exposure of this material. Similarly, the energy-independent 15\% QF adopted in \cite{cohcsi} seems hard to justify. For CE$\nu$NS at spallation sources \cite{ESS,cohcsi,clovers,Collar:2025sle} a  pragmatic decision is forced once the diminishing-returns ``brick wall'' at \mbox{$\approx$4 keVnr} in \mbox{Fig.\ 8} is assimilated: in upcoming work we will discuss the significant advantages of a large-mass room-temperature CsI[Na] detector -by now a well-established technique, with scope for improvements- over the smaller cryogenic CsI designs contemplated in  \cite{ESS,cohcsi,clovers}.

\section{Acknowledgements} We are indebted to Paolo Privitera for facilitating access to the $^{124}$Sb source. This work is supported by ERC Advanced Grant 101055120 (ESSCEvNS) and NSF Award PHY-2209579. A. Simon acknowledges additional support from  Marie Skłodowska-Curie Grant No 101026628 (vPESS), funded by the European Union’s Horizon 2020 research and innovation programme and Grant RYC2023-045436-I, funded by MICIU/AEI/10.13039/501100011033 and FSE+.

\bibliography{apssamp}

\end{document}